\documentstyle[aps,preprint]{revtex}

\begin{document}
\draft
\title{The untilted diffuse matter Bianchi V Universe}
\author{T. Christodoulakis, G. Kofinas and Vasilios Zarikas\thanks{%
E-mail addresses: tchris@cc.uoa.gr, \thinspace vzarikas@cc.uoa.gr}}
\address{University of Athens, Physics Department, \\
Nuclear and Particle Physics Division,\\
GR--15771 Athens, Greece}
\maketitle

\begin{abstract}
A diffuse matter filled Type V Universe is studied. The anisotropic
behaviour, the distortion caused to the CMBR and the parameter region
allowed by present cosmological bounds are examined. It is shown how the
overall sky pattern of temperature anisotropies changes under a
non-infinitesimal spatial coordinate transformation that preserves the Type
V manifest homogeneity.
\end{abstract}



\section{Introduction}

The general solution of Type V spatially homogenous cosmology in the
realistic case of untilted diffuse mater $p=g\ \rho $, \cite{suk}, \cite{mac}%
, \cite{kramer} is studied in various forms of the metric. This kind of
equation of state covers the actual matter composition of the Universe for
most of it's time evolution. It covers both the radiation era and the dust
epoch. Note also, that not only theoretical considerations (Mach's
principle) but also observational data, suggest a value close to zero for
the rotation of the Universe \cite{bjs83}, thus validating our assumption
about untilted fluid. This solution describes an open Universe, a feature
that is lately favorable \cite{turner}, and at the same time is not
trivially simple like the Kasner family.

The existence of a primordial global anisotropy in our Universe is an open
issue. The promising string theories suggest an anisotropic expansion that
compactified all but four dimensions. It is natural to expect that the
remaining dimensions unfold to one of the nine known types of spatially
homogenous models, most of which are anisotropic.

On the other hand it is well known that the smoothness of the cosmic
microwave background radiation (CMBR) suggests that the
Friedmann-Robertson-Walker (FRW) cosmological model is a fair description of
the actual present Universe. However the radiation spectra has small
fluctuations originating, in general, from pertubations produced by some
physical mechanism (inflation-topological defects) and from primordial
anisotropies and inhomogeneities. If the Universe experienced an
inflationary period, any primordial overall anisotropy may have been
smoothed but not erased. A possible realistic scenario then, could be that
we had, after the Planck regime, an anisotropic Universe that appeared more
isotropic after inflation and almost FRW during the last scattering and
afterwards. It is certainly important to know more about the {\it %
characteristics} and the {\it evolution} of this global geometric primordial
anisotropy.

The above mentioned scenario can be fully studied, only if we study the
general solutions of the nine types of spatially homogeneous cosmologies.
Interesting works have appeared in the literature regarding the connection
of the various Bianchi types with the observed cosmological data. There have
been works assuming small anisotropic pertubations to the FRW solution \cite
{bjs83}, \cite{ruffini}, \cite{silk} which are valid after last scattering
and others \cite{ms95}, \cite{bjpa}, which incorporate the exact solutions
for various types of homogeneous models. All these works study the final
temperature patterns induced on the CMBR. The basic conclusion is that, if
someone wants the distortion of the null geodesics after last scattering to
be small in order to comply with the observed quadrupole, then one has to
fix the parameters of the various models to produce a metric very close to
the FRW. Consequently, they set upper limits on the relative shear $\left( 
\frac{\Sigma }{\Theta }\right) _{0}$ , where $\Sigma $ is the shear
anisotropy and $\Theta =3H$ is the expansion rate, and on the vorticity.

There is also the important work of W. R. Stoeger, R. Maartens, and G. F. R.
Ellis \cite{mes95}, \cite{sme95} in which a theorem is proven stating that
if all observers measured CMBR to be almost isotropic, in our past light
cone from the last scattering till now, then this universe has to be almost
FRW in this region. They are referring to an expanding Universe with
noninteracting matter and radiation. 
Then based on this theorem and some not easily testable simplifications they
set upper limits on the allowed relative shear $\left( \frac{\Sigma }{\Theta 
}\right) _{0}$ .

One can in principle use the CMBR data in order to determine the allowed
parameter space and the type of the various homogenous cosmological
solutions. Assuming that the actual Universe after last scattering is a dust
Bianchi V one, we studied the distortion of the initial isotropic
temperature pattern caused by the null geodesics. We present a method that
yields simple system of equations for the null geodesics (a similar idea has
been used in \cite{geo}). We found that for some parameters that make the
solution close to the FRW one, it is possible to have for the period from
last scattering till now very small quadrupole anisotropy compatible with
the present data. For parameters not close to the FRW solution we always
failed to find acceptably small quadrupole anisotropy. The most important
feature we found is that a non-infinitesimal and not orthogonal
transformation of coordinates that preserves the Bianchi V manifest
homogeneity of the spacetime, leads to a non trivial transformation of the
repeatedly appearing in the literature overall sky-pattern of anisotropies,
although does not affect the covariant properties carried in that pattern.

The paper is organized as follows: In section II the Einstein equations are
setup and completely solved. We present the solution in the off diagonal
form, in order to make manifest the dependency on the spatial coordinate
system, of the calculations leading to the temperature fluctuations as a
function of the sky angles. In section III the anisotropic behaviour and its
effect on the temperature pattern, meaning the plot of the temperature as a
function of the sky angles, is studied. The magnitude of the pattern
distortion resulted by the use of the non-diagonal form of the metric is
presented. Some concluding remarks are also given.

\section{The untilted diffuse matter Bianchi V cosmology}

Einstein equations are written as 
\begin{equation}
^{(4)}R_{AB}-\frac{1}{2}\ ^{(4)}R\ g_{AB}=\kappa T_{AB},  \label{einstein}
\end{equation}
where $\kappa =-8\pi G$ the Einstein constant and $T_{AB}$ the
energy-momentum tensor. Capital Latin indices are spacetime indices $%
A,B,..=0,...,3$ while lower Latin are spatial indices $i,j,..=1,2,3.$

In the 3+1 analysis, we use the lapse and shift functions in order to write
the 4-metric as 
\begin{equation}
ds^{2}=(N_{i}N^{i}-N^{2})dt^{2}+2N_{i}dx^{i}dt+g_{ij}dx^{i}dx^{j}.
\label{le}
\end{equation}
Then, Einstein equations (\ref{einstein}) are equivalent to the following
system : 
\begin{eqnarray}
K_{i|j}^{j}-K_{|i} &=&-\kappa NT_{i}^{0}  \label{e1} \\
K_{j}^{i}K_{i}^{j}-K^{2}+^{(3)}R &=&-2\kappa \left(
T_{0}^{0}-T_{i}^{0}N^{i}\right)  \label{e2}
\end{eqnarray}
\begin{eqnarray}
&&\stackrel{.}{K}_{j}^{i}-NKK_{j}^{i}+N\
^{(3)}R_{j}^{i}+g^{il}N_{|jl}-\left(
K_{j|l}^{i}N^{l}+K_{l}^{i}N_{|j}^{l}-K_{j}^{l}N_{|l}^{i}\right)  \nonumber \\
&=&\kappa N\left( T_{j}^{i}+T_{j}^{0}N^{i}\right) -\frac{N}{2}(\kappa \
^{(4)}T)\delta _{j}^{i}  \label{e3}
\end{eqnarray}
The extrinsic curvature $K_{ij}$ is given by 
\begin{equation}
K_{ij}=\frac{1}{2N}\left( N_{i|j}+N_{j|i}-\frac{\partial g_{ij}}{\partial t}%
\right) ,
\end{equation}
while 
\begin{equation}
K_{j}^{i}\equiv g^{il}K_{lj},\qquad K\equiv K_{i}^{i}\ ,
\end{equation}
where $g^{ij}$ is the inverse of the 3-metric $g_{ij}.$ We also define 
\begin{equation}
^{(3)}R_{j}^{i}\equiv g^{il}\ ^{(3)}R_{lj},\qquad T_{B}^{A}\equiv
g^{AC}T_{CB},\qquad ^{(4)}T\equiv T_{A}^{A}\ .
\end{equation}
The class of spatially homogeneous spacetimes is characterised by the
existence of an $m$-dimensional isometry group of motions, acting on each
surface of simultaneity $\Sigma _{t}$ . If $m>3$ and there is no proper
invariant subgroup of dimension $3$, the spacetime is of the Kantowski-Sachs
type \cite{ks66}, and will not concern us further. When $m=3,$ the dimension
of $\Sigma _{t}$, there exist three basis one-forms $\sigma ^{a}$ satisfying

\begin{equation}
d\sigma ^{\alpha }=C_{\beta \gamma }^{\alpha }{}\sigma ^{\beta }\wedge
\sigma ^{\gamma }\Leftrightarrow \sigma _{i,j}^{\alpha }-\sigma
_{j,i}^{\alpha }=2C_{\beta \gamma }^{\alpha }\sigma _{i}^{\gamma }\sigma
_{j}^{\beta },  \label{rep}
\end{equation}
where $C_{\beta \gamma }^{\alpha }$ are the structure constants of the
corresponding isometry group. Greek indices number the different basis
1-forms and take values in the range $1,...,3$ . In this case there are
coordinates $t,\ x^{i}$ such that the line element \ref{le} takes the form 
\begin{equation}
ds^{2}=\left[ N_{\alpha }(t)N^{\alpha }(t)-N^{2}(t)\right] dt^{2}+2N_{\alpha
}(t)\sigma _{i}^{\alpha }(x)dx^{i}dt+\gamma _{\alpha \beta }(t)\sigma
_{i}^{\alpha }(x)\sigma _{j}^{\beta }(x)dx^{i}dx^{j}.  \label{line}
\end{equation}
Inserting the line element Eq.(\ref{line}) into the system of equations Eq.(%
\ref{e1}), Eq.(\ref{e2}), Eq.(\ref{e3}) we get, via repeated use of Eq.(\ref
{rep}) the following set of ordinary differential equations for the
Bianchi-type spatially homogeneous spacetimes: 
\begin{equation}
K_{\nu }^{\mu }C_{\alpha \mu }^{\nu }-K_{\alpha }^{\mu }C_{\mu \nu }^{\nu }=%
\frac{\kappa N}{2}T_{\alpha }^{0},\qquad  \label{eq1}
\end{equation}
\begin{equation}
K_{\beta }^{\alpha }K_{\alpha }^{\beta }-K^{2}+^{(3)}R+2\Lambda =-2\kappa
\left( T_{0}^{0}-T_{\rho }^{0}N^{\rho }\right)  \label{eq2}
\end{equation}
\begin{equation}
\stackrel{.}{K}_{\beta }^{\alpha }-NKK_{\beta }^{\alpha }+NR_{\beta
}^{\alpha }+2N^{\rho }\left( K_{\mu }^{\alpha }C_{\beta \rho }^{\mu
}-K_{\beta }^{\mu }C_{\mu \rho }^{\alpha }\right) =\kappa N\left( T_{\beta
}^{\alpha }+T_{\beta }^{0}N^{\alpha }\right) -\frac{N}{2}\left[ \kappa
(T_{0}^{0}+T_{\rho }^{\rho })\right] \delta _{\beta }^{\alpha }  \label{eq3}
\end{equation}
where 
\begin{eqnarray}
K_{\beta }^{\alpha } &\equiv &\gamma ^{\alpha \rho }K_{\rho \beta },\qquad
R_{\beta }^{\alpha }\equiv \gamma ^{\alpha \rho }R_{\rho \beta }\qquad \text{%
and} \\
K_{\alpha \beta } &=&-\frac{1}{2N}\left( \stackrel{.}{\gamma }_{\alpha \beta
}+2\gamma _{\alpha \nu }C_{\beta \rho }^{\nu }N^{\rho }+2\gamma _{\beta \nu
}C_{\alpha \rho }^{\nu }N^{\rho }\right) ,\qquad  \label{kk} \\
R_{\alpha \beta } &=&C_{\sigma \tau }^{\kappa }C_{\mu \nu }^{\lambda }\gamma
_{\alpha \kappa }\gamma _{\beta \lambda }\gamma ^{\sigma \nu }\gamma ^{\tau
\mu }+2C_{\alpha \kappa }^{\lambda }C_{\beta \lambda }^{\kappa }+2C_{\alpha
\kappa }^{\mu }C_{\beta \lambda }^{\nu }\gamma _{\mu \nu }\gamma ^{\kappa
\lambda }+  \nonumber \\
&&2C_{\beta \kappa }^{\lambda }C_{\mu \nu }^{\mu }\gamma _{\alpha \lambda
}\gamma ^{\kappa \nu }+2C_{\alpha \kappa }^{\lambda }C_{\mu \nu }^{\mu
}\gamma _{\beta \lambda }\gamma ^{\kappa \nu }.  \label{ricci}
\end{eqnarray}
The components of the energy-momentum tensor appearing in the system of Eq.(%
\ref{eq1}), Eq.(\ref{eq2}), Eq.(\ref{eq3}), are the time-dependent parts of
the corresponding full components $T_{B}^{A}$. The energy-momentum tensor of
a perfect fluid can be written as 
\begin{equation}
T^{AB}=(\rho +p)u^{A}u^{B}+pg^{AB},\qquad u^{A}u_{A}=-1,
\end{equation}
where $u^{A}$ is its four-velocity $(u_{A}\equiv g_{AB}u^{B})$, $\rho $ is
the total energy density and $p$ the pressure. The conservation equations $%
T^{AB}\,_{;B}=0$ are equivalent to 
\begin{eqnarray}
\stackrel{.}{\rho }+\Theta (\rho +p) &=&0  \label{st1} \\
(\rho +p)\stackrel{.}{u}^{A}+p_{,B}\ g^{AB}+\stackrel{.}{p}u^{A} &=&0,
\label{st2}
\end{eqnarray}
where $\Theta =-K=u_{;A}^{A}$ is the expansion scalar and the dot denotes $%
\bigtriangledown _{u}$. Now the last equations and the system Eq. (\ref{e1}%
), Eq.(\ref{e2}), Eq.(\ref{e3}) are integrable given an equation of state $%
p=p(\rho ).$ A very general equation of state is that of the diffuse matter 
\begin{equation}
p=g\rho ,\qquad \left| g\right| \leq 1  \label{eqst}
\end{equation}
with $g$ a constant. It includes the dust case for $g=0,$ gravitational
waves $(g=1/3),$ cosmic string networks $(g=-1/3),$ domain walls $(g=-2/3)$
and free massless scalar fields $(g=+1)$.

In spatially homogeneous spacetimes, if the perfect fluid moves orthogonal
to the spatial slices, we have

\begin{equation}
u^{A}=\frac{1}{N}\left( 
\begin{array}{ll}
1 & 0
\end{array}
\right)
\end{equation}
in the so called sychronous coordinate frame $(N^{a}=0)$ which in our case
it is also commoving. Using now Eq.(\ref{kk}) we can write Eq.(\ref{st1}) as
follows : 
\begin{equation}
\frac{\stackrel{.}{\rho }}{\rho +p}+\frac{\stackrel{.}{\gamma }}{2\gamma }=0,
\end{equation}
where $\gamma \equiv \det \left( \gamma _{\alpha \beta }\right) .$ The dot
denotes ordinary differentiation with respect to the time label that
characterises the slices $\Sigma _{t}$ . Integrating the above expression
with the help of Eq. (\ref{eqst}) we find 
\begin{equation}
\rho =\rho _{1}\gamma ^{-\frac{g+1}{2}}.
\end{equation}
The quantity $\rho _{1}$ is a positive constant. Note that Eq.(\ref{st2})
becomes an identity since $p=p(t)$ and $u^{A}$ is a geodesic field. The fact
that our perfect fluid is taken to be untilted leads to the simplification $%
T_{\alpha }^{0}=0$ and thus the linear equations Eq.(\ref{eq1}) coincide
with the vacuum ones, i.e. they are (when $N^{\alpha }$ vanishes) 
\begin{equation}
\gamma ^{\mu \beta }C_{\alpha \mu }^{\nu }\stackrel{.}{\gamma }_{\nu \beta
}-\gamma ^{\mu \beta }C_{\mu \nu }^{\nu }\stackrel{.}{\gamma }_{\alpha \beta
}=0.  \label{eqsgamma}
\end{equation}

We now {\em specialise} to the case of the Bianchi type V, i.e. we set $%
C_{13}^{1}=-C_{31}^{1}=C_{23}^{2}=-C_{32}^{2}=1/2$ , all other vanish. It
can be shown with the aid of a time independent automorphism inducing
diffeomorphism (A.I.D.) \cite{ckkp}, that we can choose, keeping the full
generality, $\gamma _{13}=\gamma _{23}=0,$ $\gamma _{11}\gamma _{22}-(\gamma
_{12})^{2}=(\gamma _{33})^{2}.$ It can further be shown that, after
completely integrating Einstein equations at this stage, the off diagonal
element can be rotated away through a transformation $\overline{\gamma }%
_{\alpha \beta }=\Lambda _{\alpha }^{\mu }\Lambda _{\beta }^{\nu }\ \gamma
_{\mu \nu }$ with $\Lambda $ a constant automorphism matrix of the form 
\begin{equation}
\Lambda =\left( 
\begin{array}{lll}
r & s & 0 \\ 
t & u & 0 \\ 
0 & 0 & 1
\end{array}
\right) .
\end{equation}
Such a transformation emerges as the effect on $\gamma _{\alpha \beta }$ of
a time independent ''frozen'' A.I.D., see Appendix of \cite{kofi}; Thus the
appearance of a non vanishing $\gamma _{12}$ is tantamount to the use of a
spatial coordinate system different from the system supporting the diagonal
form of the solutions. In what follows we keep the non diagonal element non
zero in order to solve the geodesic equations in this different spatial
coordinate choice and thus be able to discuss the dependency of the
temperature pattern on this choice. One would wonder why we should make the
endeavour to solve the Einstein equations in this off diagonal form and not
just take the diagonal solution and obtain a non diagonal form through an
arbitrary time dependent automorphism. The reason is that only the constant
automorphism of the form discussed above, keep us within the space of
solutions, as it can be proven using the off diagonal solution presented
here. This off diagonal solution represents the full space of solutions
(with $\gamma _{13}=\gamma _{23}=0$ of course). The metric $\gamma _{\alpha
\beta }$ is therefore taken to be 
\begin{equation}
\gamma _{\alpha \beta }=\left( 
\begin{array}{lll}
a & b & 0 \\ 
b & c & 0 \\ 
0 & 0 & f
\end{array}
\right) ,\qquad \gamma =f^{3}\Leftrightarrow ac-b^{2}=f^{2}.  \label{holo}
\end{equation}
and Eqs. (\ref{eqsgamma}) are then identically satisfied. An off-diagonal
matrix of the above form gives the following Ricci tensor Eq.(\ref{ricci})
and Ricci scalar : 
\begin{equation}
R_{\alpha \beta }=2\gamma ^{-1/3}\gamma _{\alpha \beta },\qquad
^{(3)}R=6\gamma ^{-1/3}.  \label{R}
\end{equation}
As far as time is concerned , we adopt the gauge fixing condition $N=\sqrt{%
\gamma }$ , since this simplifies the form of the equations. Consequently,
the remaining Einstein equations Eq.(\ref{eq2}), Eq.(\ref{eq3}) read 
\begin{eqnarray}
\stackrel{.}{\gamma }^{\alpha \beta }\stackrel{.}{\gamma }_{\alpha \beta
}+\left( \frac{\stackrel{.}{\gamma }}{\gamma }\right) ^{2}-24\gamma
^{2/3}+8\gamma \kappa \rho  &=&0  \label{mon} \\
\stackrel{..}{\gamma }_{\alpha \beta }-\gamma ^{\mu \nu }\stackrel{.}{\gamma 
}_{\alpha \mu }\stackrel{.}{\gamma }_{\beta \nu }+G(\gamma )\gamma _{\alpha
\beta } &=&0  \label{digr}
\end{eqnarray}
with 
\begin{equation}
G(\gamma )\equiv -4\gamma ^{2/3}+\gamma \kappa \left( \rho -p\right) .
\end{equation}
Taking the trace of Eq.(\ref{digr}), one arrives at 
\begin{equation}
\left( \frac{\stackrel{.}{\gamma }}{\gamma }\right) ^{.}+3G(\gamma )=0
\label{tr}
\end{equation}
which has a first integral 
\begin{equation}
w\equiv \left( \frac{\stackrel{.}{\gamma }}{\gamma }\right) ^{2}+U(\gamma
)=const,  \label{deq4}
\end{equation}
where 
\begin{equation}
U(\gamma )\equiv 6\int \frac{G(\gamma )}{\gamma }d\gamma .
\end{equation}
We make the conformal transformation 
\begin{equation}
\varpi _{\alpha \beta }\equiv \gamma ^{-1/3}\gamma _{\alpha \beta }
\end{equation}
and then, due to Eq.(\ref{R}) and Eq.(\ref{tr}) we can write Eq.(\ref{digr})
as 
\begin{equation}
\stackrel{..}{\varpi }_{\alpha \beta }-\varpi ^{\mu \nu }\stackrel{.}{\varpi 
}_{\alpha \mu }\stackrel{.}{\varpi }_{\beta \nu }=0.
\end{equation}
The above equation can be integrated immediately , giving 
\begin{equation}
\stackrel{.}{\varpi }_{\alpha \beta }=\vartheta _{\alpha }^{\delta }\varpi
_{\delta \beta }.  \label{www}
\end{equation}
Since $\varpi \equiv \det (\varpi _{\alpha \beta })=1\Rightarrow \vartheta
_{\alpha }^{\alpha }=0$ . The system of Eq.(\ref{www}) is equivalent to the
following set of equations 
\begin{eqnarray}
(\gamma ^{-1/3}a)^{.} &=&\vartheta _{1}^{1}(\gamma ^{-1/3}a)+\vartheta
_{1}^{2}(\gamma ^{-1/3}b)  \label{th1} \\
(\gamma ^{-1/3}b)^{.} &=&\vartheta _{1}^{1}(\gamma ^{-1/3}b)+\vartheta
_{1}^{2}(\gamma ^{-1/3}c)  \label{th2} \\
(\gamma ^{-1/3}b)^{.} &=&\vartheta _{2}^{1}(\gamma ^{-1/3}a)+\vartheta
_{2}^{2}(\gamma ^{-1/3}b)  \label{th3} \\
(\gamma ^{-1/3}c)^{.} &=&\vartheta _{2}^{1}(\gamma ^{-1/3}b)+\vartheta
_{2}^{2}(\gamma ^{-1/3}c)  \label{th4}
\end{eqnarray}
and also the relations $\vartheta _{1}^{3}=\vartheta _{2}^{3}=\vartheta
_{3}^{1}=\vartheta _{3}^{2}=\vartheta _{3}^{3}=0.$ Then $\vartheta
_{2}^{2}=-\vartheta _{1}^{1}$ . At this stage we could invoke a constant
automorphism matrix of the form earlier given and set the matrix $\vartheta
_{\beta }^{\alpha }$ to the form 
\begin{equation}
\vartheta =\left( 
\begin{array}{lll}
\vartheta _{1}^{1} & 0 & 0 \\ 
0 & -\vartheta _{1}^{1} & 0 \\ 
0 & 0 & 0
\end{array}
\right) 
\end{equation}
which would directly led us to the diagonal form of the solution \cite{suk}, 
\cite{kramer}, \cite{mac} (or Joseph's \cite{jos}, in the vacuum case).
However since our aim is to uncover the dependency of the temperature
pattern on the spatial coordinate system, we choose to keep the non
essential parameters in the game. The quadratic Einstein equation Eq.(\ref
{mon}) and the system Eq.(\ref{th1}) - Eq.(\ref{th4}) imply 
\begin{equation}
(\vartheta _{1}^{1})^{2}+\vartheta _{1}^{2}\vartheta _{2}^{1}=\frac{w}{3}-%
\frac{1}{3}U(\gamma )-12\gamma ^{2/3}+4\gamma \kappa \rho ,  \label{prix}
\end{equation}
where the function $U(\gamma )$ for the diffuse matter equation of state Eq.(%
\ref{eqst}) is written as

\begin{equation}
U(\gamma )=-36\gamma ^{2/3}+12\kappa \rho _{1}\gamma ^{\frac{1-g}{2}}.
\end{equation}
Then Eq.(\ref{prix}) reduces the independent constants : 
\begin{equation}
\frac{w}{3}=(\vartheta _{1}^{1})^{2}+\vartheta _{1}^{2}\vartheta _{2}^{1}.
\label{con}
\end{equation}
For $b\neq 0$ , $\vartheta _{2}^{1}\vartheta _{1}^{2}\neq 0,$ we have from
Eq.(\ref{th2}), Eq.(\ref{th3}) and Eq.(\ref{holo}) that 
\begin{equation}
\vartheta _{2}^{1}a^{2}-2\vartheta _{1}^{1}ab-\vartheta
_{1}^{2}b^{2}=\vartheta _{1}^{2}\gamma ^{2/3}.  \label{fin}
\end{equation}
Since Eq.(\ref{th4}) can be derived from the other three of the system Eq.(%
\ref{th1}) - Eq.(\ref{th4}), it suffices to solve the system of Eq. (\ref
{th1}), Eq.(\ref{th3}) and Eq.(\ref{fin}). The solution of the system Eq. (%
\ref{th1}), Eq.(\ref{th3}) gives the expressions $\gamma ^{-1/3}a$ , $\gamma
^{-1/3}b$ as sums of two exponential functions of time. Only for $w>0$ these
solutions satisfy also Eq.(\ref{fin}). Thus finally, the general solution of
the Bianchi V universe with an orthogonal perfect fluid and linear equation
of state, in the off diagonal form, is given as follows : 
\begin{eqnarray}
a(t) &=&\gamma ^{1/3}(t)\left[ k\exp \left( \sqrt{\frac{w}{3}}\ t\right) +%
\frac{3(\vartheta _{1}^{2})^{2}}{4wk}\exp \left( -\sqrt{\frac{w}{3}}\
t\right) \right]   \nonumber \\
b(t) &=&\gamma ^{1/3}(t)\left[ \frac{k}{\vartheta _{1}^{2}}\left( \sqrt{%
\frac{w}{3}}-\vartheta _{1}^{1}\right) \exp \left( \sqrt{\frac{w}{3}}\
t\right) -\frac{3\vartheta _{1}^{2}}{4wk}\left( \sqrt{\frac{w}{3}}+\vartheta
_{1}^{1}\right) \exp \left( -\sqrt{\frac{w}{3}}\ t\right) \right]   \nonumber
\\
c(t) &=&\gamma ^{1/3}(t)\left[ \frac{k}{(\vartheta _{1}^{2})^{2}}\left( 
\sqrt{\frac{w}{3}}-\vartheta _{1}^{1}\right) ^{2}\exp \left( \sqrt{\frac{w}{3%
}}\ t\right) +\frac{3}{4wk}\left( \sqrt{\frac{w}{3}}+\vartheta
_{1}^{1}\right) ^{2}\exp \left( -\sqrt{\frac{w}{3}}\ t\right) \right]  
\nonumber \\
f(t) &=&\gamma ^{1/3}(t)=\left[ h^{-1}(\frac{t}{6})\right] ^{2},\qquad
h(x)=\int \frac{dx}{x\sqrt{w+36x^{4}-12\kappa \rho _{1}x^{3}}}\ .
\label{solution}
\end{eqnarray}
The solution contains five arbitrary constants $k,$ $\vartheta _{1}^{1},$ $%
\vartheta _{1}^{2},\vartheta _{2}^{1}$ $,$ $\rho _{1}$ $.$ The integration
constant that arises in the expression giving $\gamma (t)$ is absorbable
(non-essential) and is fixed by choosing the zero point on the time axis.
One can further realise that three of the five constants are also absorbable
as explained above, leaving us with two essential parameters one for the
matter and one for the gravity, as expected. The counting of the number of
essential parameters for the vacuum Bianchi cosmologies can be understood as
follows: From the initial twelve constants $\gamma _{\alpha \beta }(t),%
\stackrel{\cdot }{\gamma }_{\alpha \beta }(t_{o})$ we subtract twice the
number of independent first class constraints and the number of outer
automorphism parameters. In our Type V case this reach $12-2\times 4-3=1$
essential parameter. For Type II we have $12-2\times 3-4=2$ and likewise for
other Types.

If one employs the above derived solution to describe the dust filled
Universe, after last scattering, it is possible to constraint some of the
free parameters. Thus 
\begin{equation}
w=\frac{144}{H^{4}}\frac{q-\frac{\Omega }{2}}{\left( 2-q-\frac{3\Omega }{2}%
\right) ^{3}},
\end{equation}
\begin{equation}
\rho _{1}=\frac{-1}{\kappa }\frac{3\sqrt{8}\Omega }{H\left( 2-q-\frac{%
3\Omega }{2}\right) ^{3/2}},
\end{equation}
\begin{equation}
\gamma =\frac{1}{H^{6}}\frac{8}{\left( 2-q-\frac{3\Omega }{2}\right) ^{3}}\ ,
\label{go}
\end{equation}
where $q$ is the decceleration parameter, $H$ is the Hubble rate and $\Omega 
$ is the flatness parameter. In their definition the mean scale factor $%
\gamma $ , has been used. The above expressions will be used to calculate
the constants $w,$ $\rho _{1},$ $\gamma _{today}=\gamma _{o}$ from the
measurable $H_{o},q_{o},$ $\Omega _{o}.$

The above relations impose the following constraints on the allowed
parameter space of the solution: 
\begin{equation}
2-q-\frac{3\Omega }{2}>0,\qquad q-\frac{\Omega }{2}>0.
\end{equation}
The above constraints imply that $0<q<2$ and $0<\Omega <1.$

\section{The anisotropic behavior}

It is well known \cite{ch73} that, under the assumptions of zero
cosmological constant, the validity of the dominant energy condition and the
positiveness of pressure, Bianchi types I, V and VII are the only ones that
can satisfy four isotropisation conditions enlisted below. Of course this by
no means implies that the solution will indeed isotropise. We have to know
the exact solutions in order to examine the anisotropic behaviour. These
four conditions of C. B. Collins, S. W. Hawking, are as follows :

\begin{enumerate}
\begin{description}
\item  (i)\qquad $\gamma \rightarrow \infty $

\item  (ii)\qquad $T^{00}>0$ and $\frac{T^{0i}}{T^{00}}\rightarrow 0$

\item  (iii)\qquad $\frac{\Sigma }{\Theta }\rightarrow 0$ , where $2\Sigma
^{2}\equiv \sigma ^{AB}\sigma _{AB}$

\item  (iv)\qquad $\gamma ^{-1/3}\gamma _{\alpha \beta }\rightarrow const$
\end{description}
\end{enumerate}

The limits are understood to be taken as the proper time approach infinity $%
\tau \rightarrow +\infty $. Referring to the first condition let us study
the behaviour of the quantity $\gamma (\tau )$, for the metric presented in
the previous section. The proper time $\tau $ is given, in the dust case and
for an expanding universe, by the following relation 
\begin{equation}
\tau (x)=\int_{0}^{x}F(x)dx,\qquad F(x)\equiv \frac{6x^{2}}{\sqrt{w-12\kappa
\rho _{1}x^{3}+36x^{4}}},
\end{equation}
where $x\equiv \gamma ^{1/6}$ and the choice of the origin of time is $\tau
=0$ for $\gamma =0.$ It is easily verified that $\gamma \rightarrow \infty $
for $\tau \rightarrow +\infty .$ The second condition is trivially satisfied
due to the orthogonality of the four velocity $u^{A}.$ For the third
criterion, we can compute the shear anisotropy $\Sigma $ via the shear
tensor $\sigma ^{AB}$ in the basis where $N^{\alpha }=0$%
\begin{equation}
2\Sigma ^{2}=K_{\beta }^{\alpha }K_{\alpha }^{\beta }-\frac{1}{3}K^{2}=\frac{%
1}{6N^{2}}\left( \frac{\stackrel{.}{\gamma }}{\gamma }\right) ^{2}-\left(
^{(3)}R+2\kappa T_{0}^{0}\right) .
\end{equation}
Finally, we find for the solution Eq.(\ref{solution}) 
\begin{equation}
\frac{\Sigma }{\Theta }=\left[ \frac{1}{3}\frac{w}{w-U(\gamma )}\right]
^{1/2}.
\end{equation}
One should note that for $\gamma \rightarrow 0$ we have $\frac{\Sigma }{%
\Theta }\rightarrow \frac{1}{\sqrt{3}}.$ Furthermore, the function $\frac{%
\Sigma }{\Theta }$ is a strictly decreasing function of $\gamma $ and $%
\lim_{\gamma \rightarrow \infty }\frac{\Sigma }{\Theta }=0.$ The last
condition (iv), holds for dust, since we can prove that $\lim_{x\rightarrow
+\infty }t(x)<+\infty $ . Therefore the solution satisfies all the specified
conditions. In the following, we will examine some more criteria, that can
characterise the behaviour of a homogeneous model.

A cosmological model with a perfect fluid source is said, according to GFR
Ellis \cite{ellis}, to be close to a FRW model in some open set of the
spacetime, if for some suitably small constant $\varepsilon \ll 1$ the
following inequalities hold, in this open set: 
\begin{equation}
\frac{\Sigma }{\Theta }<\varepsilon ,\qquad \frac{\sqrt{\left| E^{\mu \nu
}E_{\mu \nu }\right| }}{H^{2}}<\varepsilon ,\qquad \frac{\sqrt{\left| H^{\mu
\nu }H_{\mu \nu }\right| }}{H^{2}}<\varepsilon ,  \label{elliscriteria}
\end{equation}
where $E_{\lambda \nu }\equiv C_{\lambda \mu \nu \kappa }\ \eta ^{\mu }\eta
^{\kappa },$ $H_{\lambda \nu }\equiv \widetilde{C}_{\lambda \mu \nu \kappa
}\ \eta ^{\mu }\eta ^{\kappa }$ are the electric and magnetic parts of the
Weyl tensor $C_{\lambda \mu \nu \kappa }$ respectively. The dual Weyl tensor 
$\widetilde{C}_{\lambda \mu \nu \kappa }$ is defined as $\widetilde{C}%
_{\lambda \mu \nu \kappa }\equiv \frac{1}{2}\eta _{ab}^{st}C_{stcd}$ with $%
\eta ^{abcd}$ the complete antisymmetric tensor and $\eta ^{1234}\equiv 
\frac{1}{\sqrt{-g}}$ . These covariant criteria are believed to indicate
that the model under consideration is close to a FRW one. For the solution
considered here we find that 
\begin{equation}
\frac{\sqrt{\left| E^{\mu \nu }E_{\mu \nu }\right| }}{H^{2}}=\sqrt{6w}\frac{%
\sqrt{\left| 2w+3U\right| }}{w-U},\qquad \frac{\sqrt{\left| H^{\mu \nu
}H_{\mu \nu }\right| }}{H^{2}}=6\sqrt{6w}\frac{\gamma ^{1/3}}{w-U}.
\end{equation}
These inequalities always hold in some interval $\left( \gamma _{1,}\gamma
_{2}\right) $ if we choose a sufficient small value for the parameter $w.$

Finally the distortion that the geometric anisotropies induce to the
temperature pattern via the null geodesics will be examined. Through this
analysis it will be possible to find also the parameter region that is
allowed from the current CMBR constraints. The small anisotropies of the
CMBR are studied computing the angular variation of the microwave
temperature, $T_{o}(\theta ,\varphi )$ characterising the frequency
distribution of the radiation. One defines 
\begin{equation}
\frac{\Delta T}{T}(\theta ,\varphi )\equiv \frac{T_{o}(\theta ,\varphi
)-\left\langle T_{o}\right\rangle }{\left\langle T_{o}\right\rangle },
\end{equation}
where $\left\langle T_{o}\right\rangle $ is the mean, over the sky, observed
temperature. The above field is decomposed in spherical harmonics 
\begin{equation}
\frac{\Delta T}{T}(\theta ,\varphi )=\sum_{l=1}^{\infty
}\sum_{m=-l}^{l}a_{lm}Y_{lm}(\theta ,\varphi ).
\end{equation}
The various multipole moments $a_{lm}$ are defined by 
\begin{equation}
a_{lm}=\int_{0}^{2\pi }d\varphi \int_{0}^{\pi }d\theta \sin \theta \ \frac{%
\Delta T}{T}Y_{l,-m}(\theta ,\varphi )
\end{equation}
and we can finally define 
\begin{equation}
(a_{l})^{2}\equiv \left\langle \left| a_{lm}\right| ^{2}\right\rangle =\frac{%
1}{4\pi }\sum_{m=-l}^{l}\left| a_{lm}\right| ^{2}.
\end{equation}
After subtracting the dipole moment of anisotropy which is ascribed to the
local motion of our sun and our galaxy around the local supercluster, one
finds a small anisotropy of the order of $10^{-5}$. Part of the remaining
anisotropy is the quadrupole fluctuation.

The cosmological redshift $z_{e}$ of a photon emitted at last scattering can
be expressed as \cite{e71} 
\begin{equation}
1+z_{e}=\frac{\left( u_{A}k^{A}\right) _{e}}{\left( u_{A}k^{A}\right) _{o}},
\label{z}
\end{equation}
where $k^{A}$ is the tangent vector to the null geodesics and $u^{A}$ is the
four velocity which in our case is vertical to the homogenous slices. The
subscripts denote the emission time and the present time. It is obvious that 
\begin{equation}
u_{A}k^{A}=-N(t)k^{0}.  \label{k0}
\end{equation}
A unit vector, tangent to the spatial homogenous slice today $t_{o}$, in our
position $O$, can be written as 
\begin{equation}
e^{A}\equiv \left( u^{A}\right) _{o}+\frac{\left( k^{A}\right) _{o}}{\left(
u_{A}k^{A}\right) _{o}}=\sin \theta \cos \varphi \ Y_{1}+\sin \theta \sin
\varphi \ Y_{2}+\cos \theta \ Y_{3},
\end{equation}
where the sky angles $\theta ,\varphi $ specify the direction of the
incoming photon. Note that we introduced on the spatial hypersurface an
orthonormal basis $\{Y_{i}\}$ with $g(Y_{i},Y_{j})=\delta _{ij}$ . It is
related to the invariant basis $\{\sigma ^{\alpha }\}$ : 
\begin{equation}
\sigma ^{1}=e^{-x^{1}}dx^{2},\qquad \sigma ^{2}=e^{-x^{1}}dx^{3},\qquad
\sigma ^{3}=dx^{1}.
\end{equation}
\begin{equation}
X_{1}=e^{-x^{1}}\partial _{2},\qquad X_{2}=e^{-x^{1}}\partial _{3},\qquad
X_{3}=\partial _{1}.
\end{equation}
by the following relations 
\[
Y_{1}=\frac{1}{\sqrt{a_{o}}}X_{1},\qquad Y_{2}=\frac{1}{\sqrt{f_{o}}}%
X_{3},\qquad Y_{3}=\frac{\sqrt{a_{o}}}{f_{o}}\left( X_{2}-\frac{b_{o}}{a_{o}}%
X_{1}\right) . 
\]
where $\partial _{i}\equiv \partial /\partial x^{i}$. The symbols $a_{o},\
b_{o},\ f_{o}$ denote the scale factors evaluated at present. We can
parametrise our position without losing the generality demanding $\left(
x^{i}\right) _{o}=0$ .Then we get 
\[
Y_{1}=\frac{1}{\sqrt{a_{o}}}\partial _{2},\qquad Y_{2}=\frac{1}{\sqrt{f_{o}}}%
\partial _{1},\qquad Y_{3}=\frac{\sqrt{a_{o}}}{f_{o}}\left( \partial _{3}-%
\frac{b_{o}}{a_{o}}\partial _{2}\right) . 
\]
Finally we can express the present components of the tangent to the null
geodesics vector from Eq.(\ref{z}), Eq.(\ref{k0}) 
\begin{eqnarray}
(k^{1})_{o} &=&-(k^{0})_{o}\sqrt{\frac{\gamma _{o}}{f_{o}}}\sin \theta \sin
\varphi \ ,  \label{3ko} \\
(k^{2})_{o} &=&-(k^{0})_{o}\sqrt{\frac{\gamma _{o}}{a_{o}}}\left( \sin
\theta \cos \varphi -\frac{b_{o}}{a_{o}}\cos \theta \right) \ ,  \nonumber \\
(k^{3})_{o} &=&-(k^{0})_{o}\frac{\sqrt{\gamma _{o}a_{o}}}{f_{o}}\cos \theta
\ .  \nonumber
\end{eqnarray}

Our purpose is to find the distortion that the anisotropy of the homogenous
spatial slices induce, in the propagation of the photons and consequently in 
$z_{e}(\theta ,\varphi )$ . It is well known that the inner product of a
geodesic field $k^{A}$ with a Killing vector field $\xi ^{A}$ is constant
along the geodesics. In our case, there are three Killing vectors tangent to
the homogenous hypersurfaces \cite{rs75}, 
\begin{equation}
\xi _{1}=\partial _{2},\qquad \xi _{2}=\partial _{3},\qquad \xi
_{3}=\partial _{1}+x^{2}\partial _{2}+x^{3}\partial _{3}
\end{equation}
The three constant quantities, along the geodesics, $K_{\alpha }=g_{AB}$ $%
\xi _{\alpha }^{A}k^{B}$ can be used to determine the null vector as 
\begin{eqnarray}
k^{1} &=&\gamma ^{-1/3}\left( K_{3}-K_{1}x^{2}-K_{2}x^{3}\right)  \label{kk1}
\\
k^{2} &=&\gamma ^{-2/3}e^{2x^{1}}\left( K_{1}\gamma _{22}-K_{2}\gamma
_{12}\right)  \label{kk2} \\
k^{3} &=&\gamma ^{-2/3}e^{2x^{1}}\left( K_{2}\gamma _{11}-K_{1}\gamma
_{12}\right) .  \label{kk3}
\end{eqnarray}
From Eq.(\ref{3ko}) we can compute the three constants 
\begin{eqnarray}
K_{\alpha } &=&-\left( k^{0}\right) _{o}\ \Lambda _{\alpha }\ ,  \label{kka}
\\
\Lambda _{1} &=&\frac{b_{o}}{c_{o}}\Lambda _{2}+\frac{\gamma _{o}^{7/6}}{%
c_{o}\sqrt{a_{o}}}\left( \sin \theta \cos \varphi -b_{o}\gamma
_{o}{}^{-1/3}\cos \theta \right) \ , \\
\Lambda _{2} &=&c_{o}\left[ \sqrt{a_{o}}\ \gamma _{o}{}^{1/6}\cos \theta +%
\sqrt{\frac{\gamma _{o}}{a_{o}}}\frac{b_{o}}{c_{o}}\left( \sin \theta \cos
\varphi -b_{o}\gamma _{o}{}^{-1/3}\cos \theta \right) \right] \ , \\
\Lambda _{3} &=&\gamma _{o}{}^{2/3}\sin \theta \sin \varphi \ .
\end{eqnarray}
If $\zeta $ is an affine parameter along the null geodesics then $%
k^{A}=dx^{A}/d\zeta $ . The last relation and $ds^{2}=0$ , give using Eq.(%
\ref{kka}), and Eqs.(\ref{kk1}-\ref{kk3}). 
\begin{equation}
k^{0}=\frac{\left( k^{0}\right) _{o}}{\sqrt{\gamma }}\Xi (\theta ,\varphi
,t_{o};t)  \label{photon}
\end{equation}
\begin{equation}
\Xi (\theta ,\varphi ,t_{o};t)\equiv \left[ \gamma ^{-1/3}\left( \Lambda
_{3}-\Lambda _{1}x^{2}-\Lambda _{2}x^{3}\right) ^{2}+e^{2x^{1}}\gamma
^{-2/3}\left( \Lambda _{1}^{2}\gamma _{22}+\Lambda _{2}^{2}\gamma
_{11}-2\Lambda _{1}\Lambda _{2}\gamma _{12}\right) \right] ^{1/2}.
\end{equation}
Finally, we find from Eq.(\ref{photon}) and Eqs.(\ref{kk1}-\ref{kk3}) that
the $x^{i}$ along the geodesics are given by 
\begin{eqnarray}
\frac{dx^{1}}{dt} &=&-\gamma (t)^{1/6}\frac{1}{\Xi (t)}\left( \Lambda
_{3}-\Lambda _{1}x^{2}-\Lambda _{2}x^{3}\right)  \label{deq1} \\
\frac{dx^{2}}{dt} &=&-\gamma (t)^{-1/6}\frac{1}{\Xi (t)}e^{2x^{1}}\left[
\Lambda _{1}\ c(t)-\Lambda _{2}\ b(t)\right]  \label{deq2} \\
\frac{dx^{3}}{dt} &=&-\gamma (t)^{-1/6}e^{2x^{1}}\left[ \Lambda _{2}\
a(t)-\Lambda _{1}\ b(t)\right] .  \label{deq3}
\end{eqnarray}
This system of equations together with the initial condition $%
x^{i}(t_{o})\equiv x_{o}^{i}=0$ , $\gamma (t_{o})=\gamma _{o}$ (see Eq. \ref
{go}) can be integrated numerically. The results of this integration will
then be used to evaluate the coordinates $x^{i}(t_{e})$ of photons, at the
emission time $t_{e}$, and then their redshift. From Eqs. (\ref{z}), (\ref
{k0}), (\ref{photon}) we find 
\begin{equation}
1+z_{e}(\theta ,\varphi )=\frac{T_{e}}{T_{o}(\theta ,\varphi )}=\gamma
_{o}^{-1/2}\ \Xi (\theta ,\varphi ,t_{o};t_{e})  \label{gauge}
\end{equation}
Note that the dependency on the parameters of the finite coordinate
transformation is manifest in the equation Eq.(\ref{gauge}). In the diagonal
form the system of equations describing the null geodesics can be obtained
from the system Eq.(\ref{deq4}), Eqs.(\ref{deq1}-\ref{deq2}-\ref{deq3})
effectively, by chopping off the terms containing $b,\ b_{o}.$ The observed
quantity $\frac{\Delta T}{T}$ is related with the above expressions through 
\begin{equation}
\frac{\Delta T}{T}(\theta ,\varphi )=\frac{T_{o}/T_{e}}{\left\langle
T_{o}/T_{e}\right\rangle }-1,
\end{equation}
where the mean value is understood as an integration over all the sky angles 
$\theta ,\varphi .$ The quadrupole moment of the temperature field $\frac{%
\Delta T}{T}$ is given by 
\begin{equation}
a_{2}=\frac{5}{16\pi ^{3/2}}\frac{1}{\left| \left\langle
T_{o}/T_{e}\right\rangle \right| }\left[ 3\left( I_{1}^{2}+I_{2}^{2}\right)
+12\left( I_{3}^{2}+I_{4}^{2}\right) +I_{5}^{2}\right] ^{1/2},
\end{equation}
where 
\begin{eqnarray}
I_{1} &\equiv &\int_{0}^{2\pi }d\varphi \int_{0}^{\pi }d\theta \sin
^{3}\theta \sin 2\varphi \frac{T_{o}}{T_{e}}(\theta ,\varphi ) \\
I_{2} &\equiv &\int_{0}^{2\pi }d\varphi \int_{0}^{\pi }d\theta \sin
^{3}\theta \cos 2\varphi \frac{T_{o}}{T_{e}}(\theta ,\varphi ) \\
I_{3} &\equiv &\int_{0}^{2\pi }d\varphi \int_{0}^{\pi }d\theta \sin
^{2}\theta \cos \theta \sin \varphi \frac{T_{o}}{T_{e}}(\theta ,\varphi ) \\
I_{4} &\equiv &\int_{0}^{2\pi }d\varphi \int_{0}^{\pi }d\theta \sin
^{2}\theta \cos \theta \cos \varphi \frac{T_{o}}{T_{e}}(\theta ,\varphi ) \\
I_{5} &\equiv &\int_{0}^{2\pi }d\varphi \int_{0}^{\pi }d\theta \sin \theta \
\left( 3\cos ^{2}\theta -1\right) \frac{T_{o}}{T_{e}}(\theta ,\varphi )\ 
\end{eqnarray}
and 
\begin{equation}
\left\langle T_{0}/T_{e}\right\rangle =\frac{1}{4\pi }\int_{0}^{2\pi
}d\varphi \int_{0}^{\pi }d\theta \sin \theta \ \frac{T_{o}}{T_{e}}(\theta
,\varphi ).
\end{equation}
The results of the numerical work done, for the general Bianchi V
cosmological solution with dust, are as follows:

\begin{itemize}
\item  The solution in the off diagonal form has five arbitrary parameters.
We determine two of the five parameters, choosing values for $q_{o},$ $%
\Omega _{o},$ $H_{o}$. Doing this we also determine the present value of the
shear anisotropy $\left( \frac{\Sigma }{\Theta }\right) _{today}\equiv
\left( \frac{\Sigma }{\Theta }\right) _{o}$ which is small if $q_{o}$ $%
\simeq \Omega _{o}/2$. We are left with three free parameters $k,$ $%
\vartheta _{1}^{1},$ $\vartheta _{1}^{2}$. These extra three parameters,
which are not present in the diagonal dust Bianchi solution, studied in \cite
{bjpa}, determine how much the diagonal elements differ inter se as well as
with the non diagonal element. The choice of different values for these
extra parameters select a solution in the space of the g.c.t. equivalent
metrics. However these different choices must not change the multipole
moments since they are true scalar numbers. Indeed the numerical calculation
exhibits exactly this; the quadrupole moment $a_{2},$ for the diagonal and
the non-diagonal form is the same up to next to next order of magnitude. We
find that the anisotropy and consequently the temperature fluctuations
induced from the geometry, {\em are small} if the shear anisotropy $\left( 
\frac{\Sigma }{\Theta }\right) _{o}$and the electric and magnetic scalars $%
\left( \frac{\sqrt{\left| E^{\mu \nu }E_{\mu \nu }\right| }}{H^{2}}\right)
_{o},$ $\left( \frac{\sqrt{\left| H^{\mu \nu }H_{\mu \nu }\right| }}{H^{2}}%
\right) _{o}$are extremely small. All these three scalars are attaining very
small values if the value of $w$ is close to zero.

\item  Compared to the previous results in \cite{bjpa} we found somewhat
smaller temperature fluctuations induced from a dust Bianchi V solution to
the CMBR for $\left( \frac{\Sigma }{\Theta }\right) _{o}=10^{-9}.$ We
computed the temperature map and we found that, both for the solution in the
diagonal form, analogous to the solution used in \cite{bjpa}\footnote{%
However the results in this work, are given for matter content, consisting
of dust and radiation.}, and for the solution written in the off diagonal
form, the temperature allover the sky can be $T_{o}=2.7277$ $\pm 10^{-5}$
when $\left( \frac{\Sigma }{\Theta }\right) _{o}=10^{-9}.$ As we have
already pointed out, this numerical value does not depend on the choice of
values for the three free parameters $k,$ $\vartheta _{1}^{1},$ $\vartheta
_{1}^{2}$ .

\item  In solving numerically the system Eq.(\ref{deq4}), Eqs.(\ref{deq1}-%
\ref{deq2}-\ref{deq3}) one has to exercise great care and demand large
accuracy\footnote{%
Large working precision is required, much more than the double precision
(16-digits) normally used.} since otherwise large anisotropies, of non
geometric origin, can appear as a result of the poor precision.

\item  The anisotropic distortion $\left( \frac{\Delta T}{T}\right) _{{\rm %
rms}}=\sqrt{\frac{1}{4\pi }\int_{0}^{2\pi }d\varphi \int_{0}^{\pi }d\theta
\sin \theta \ \left( \frac{\Delta T}{T}\right) ^{2}},$of the CMBR is smaller
for small values of the flatness parameter $\Omega _{o}$ as can be seen
below. 
\[
\begin{tabular}{||c|ccccc||}
\hline\hline
& for & $\left( \frac{\Sigma }{\Theta }\right) _{o}$ & $=10^{-9}$ &  &  \\ 
\hline\hline
$\Omega _{o}$ & $0.1$ & \multicolumn{1}{|c}{$0.2$} & \multicolumn{1}{|c}{$0.4
$} & \multicolumn{1}{|c}{$0.8$} & \multicolumn{1}{|c||}{$0.98$} \\ \hline
$\left( \frac{\Delta T}{T}\right) _{{\rm rms}}$ & $10^{-6}$ & 
\multicolumn{1}{|c}{$10^{-6}$} & \multicolumn{1}{|c}{$10^{-5}$} & 
\multicolumn{1}{|c}{$10^{-5}$} & \multicolumn{1}{|c||}{$10^{-4}$} \\ 
\hline\hline
\end{tabular}
\]

\item  We found small values of $a_{2}$ coming after integrating all the
sky. 
\[
\begin{tabular}{||c|ccccc||}
\hline\hline
& for & $\left( \frac{\Sigma }{\Theta }\right) _{o}$ & $=10^{-9}$ &  &  \\ 
\hline
$\Omega _{o}$ & $0.1$ & \multicolumn{1}{|c}{$0.2$} & \multicolumn{1}{|c}{$0.4
$} & \multicolumn{1}{|c}{$0.8$} & \multicolumn{1}{|c||}{$0.98$} \\ \hline
$\left( a_{2}\right) _{\text{all the sky}}$ & $10^{-6}$ & 
\multicolumn{1}{|c}{$5\times 10^{-6}$} & \multicolumn{1}{|c}{$10^{-5}$} & 
\multicolumn{1}{|c}{$10^{-4}$} & \multicolumn{1}{|c||}{$10^{-4}$} \\ 
\hline\hline
\end{tabular}
\]

\item  An interesting outcome of our calculations is that the solution in
the off diagonal form yields a temperature map all over the sky that is
different from the one resulting in the diagonal form. The dust diagonal
metric, we used, gives a quadrupole like structure presented in Fig. 1,
while the non diagonal metric gives a quadrupole like structure given in Fig
2. Both Figures are drawn for $\Omega _{o}=0.4$ and $\left( \frac{\Sigma }{%
\Theta }\right) _{o}=10^{-9}$ . Note that the actual observed angles in the
sky are shifted from the ones used in calculations $\theta _{obs}=\pi
-\theta ,\ \varphi _{obs}=\pi +\varphi $ .

\item  As far as the focusing effect is concerned, known to appear in open
anisotropic models\cite{nov}, the depicted situation inferred from Figure 1
is in complete agreement with the findings in \cite{bjpa}; the distance
between the focusing point and the points of the maximum anisotropy is about 
$\pi /8$. If one wants to reproduce the corresponding Figure in \cite{bjpa},
from our Figure 1, one must shift properly the axes (the focusing point
which is close to $\theta =\pi /2,\ \varphi =\pi /2$ should be shifted to $%
\theta =0,\ \varphi =0$ ). Lastly we observe that in Figure 2 one gets the
same magnitude of the focusing angle.
\end{itemize}

\section{Conclusions}

The exact solution of a Type V Bianchi Universe consisted of an untilted
perfect fluid with a diffuse matter's equation of state is studied. A simple
method for solving the null geodesics is employed, which makes use of some
appropriate inner products, constant along the geodesics.

We found that small temperature fluctuations consisted with the
observational data can be achieved if the parameters of the model are such
that give $\left( \frac{\Sigma }{\Theta }\right) _{o}\leq 10^{-9}.$ This
fact justifies our assessment that the model presented, can be a realistic
description of our Universe at least for the epoch commencing from last
scattering and reaching the present time.

The most important feature of the model studied is that the temperature
pattern produced by the geometric anisotropies, repeatedly calculated in the
literature, estimated after solving the null geodesics, is a strongly
dependent on the spatial coordinate system used, picture. We exhibited how
large this dependency is.

\acknowledgments %

We would like to thank Woei Chet for enlightening comments.

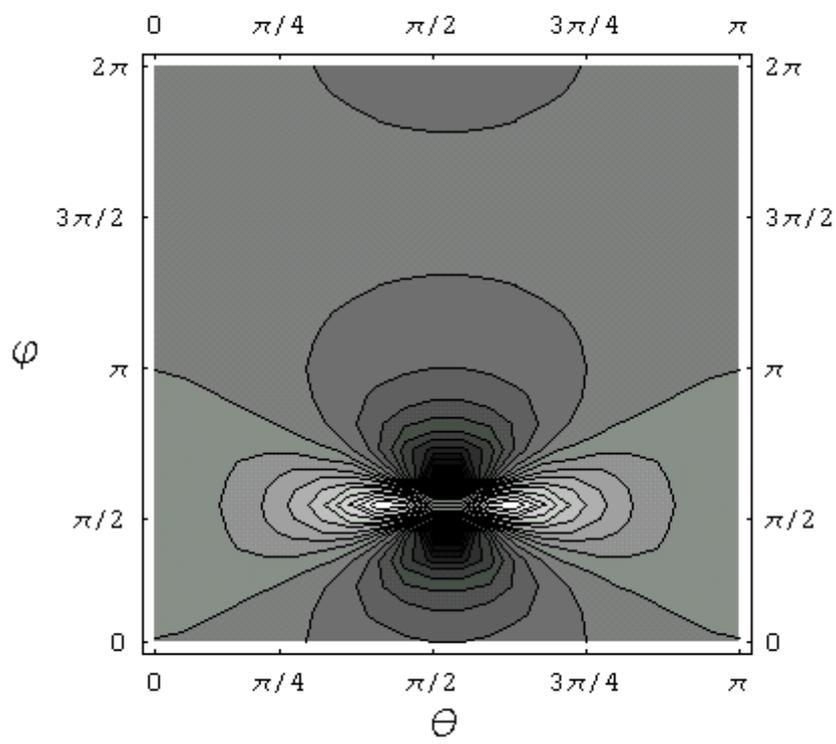
\begin{figure}[tbp]
\caption{The temperature map for the diagonal solution with $\Omega _{o}=0.4 
$. Regions with higher temperature values are lighter.}
\label{diag}
\end{figure}

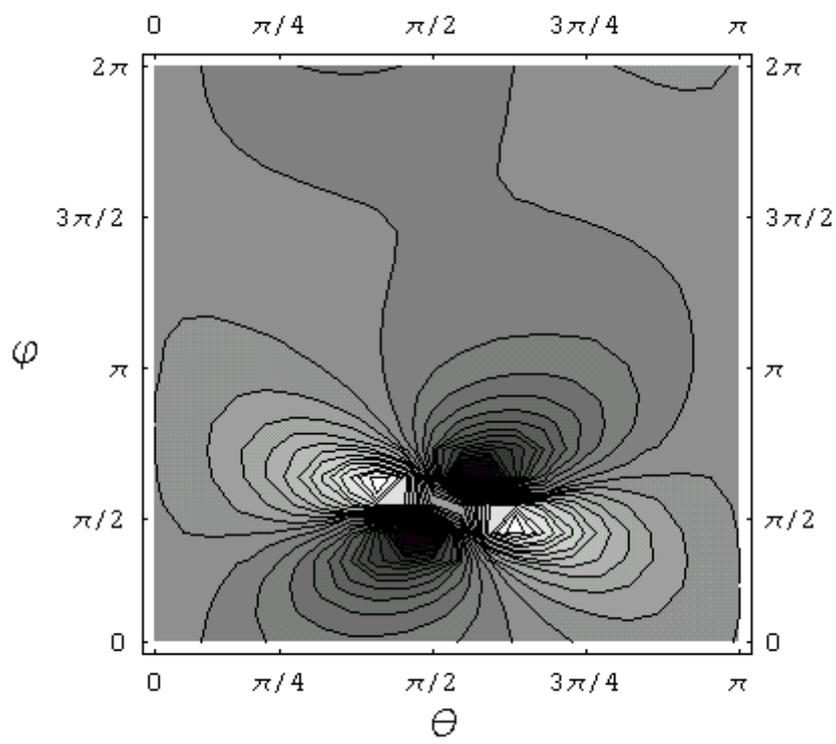
\begin{figure}[tbp]
\caption{The temperature map for the non diagonal solution with $\Omega
_{o}=0.4$.}
\label{nondiag}
\end{figure}

\end{document}